\documentclass{ws-p8-50x6-00}

\begin{document}

\title{Active-Sterile neutrino oscillations in the early Universe and the 
atmospheric neutrino anomaly}

\author{P. Di Bari}

\address{Dipartimento di Fisica, Universit\`a di Roma `La Sapienza' and INFN Roma 1,
P.le Aldo Moro, 2, I00185 Roma, Italy\\E-mail: dibari@roma1.infn.it} 

\maketitle

\abstracts{Cosmology cannot rule out the solution
$\nu_{\mu}\leftrightarrow\nu_{s}$ to the atmospheric neutrino data and thus 
only Earth experiments will be able to give a definitive answer. 
This conclusion holds when a generation of lepton number is taken into 
account and one assumes that the sterile neutrino is also slightly mixed with 
an ${\rm eV}-\tau$ neutrino. This result cannot be spoiled by a chaotic 
generation of lepton domains.}

\section{Severe BBN constraints in the two neutrino mixing scenario}

The atmospheric neutrino data are nicely explained in terms
of neutrino oscillations $\nu_{\mu}\leftrightarrow\nu_{\alpha}$. 
Even though recent results favour the solution $\alpha=\tau$, 
the possibility that $\nu_{\alpha}$ is a sterile neutrino is still 
not completely excluded by Earth experiments. In anycase it is an 
interesting issue to know whether the BBN bound is able to rule 
out the solution $\nu_{\mu}\leftrightarrow\nu_{s}$.
Neutrino oscillations are potentially able to thermalize 
the sterile neutrino that would thus contribute as a fourth
neutrino species to the expansion rate, modifying the  Standard 
BBN results. The recent indications from quasar absorbers
for low values of Deuterium abundance ($D/H\sim 3.4\times 10^{-5}$)
\cite{lowdeu}, suggest a BBN bound 
$\Delta N_{\nu}^{\rm eff}< 0.9\, (0.6)$ at 99.7 \% (95.4\%) c.l. \cite{Lisi99}. 
In this case the solution $\nu_{\mu}\leftrightarrow \nu_{s}$ 
to the atmospheric neutrino data is definitely ruled out, 
as it is shown in the left panel of figure 1. 
In this figure the constraints 
in the $\sin^{2}2\theta_{0}-|\delta m^{2}|$ plane
\begin{figure}[t]
\epsfxsize=11pc 
\vspace{-3mm}
\epsfbox{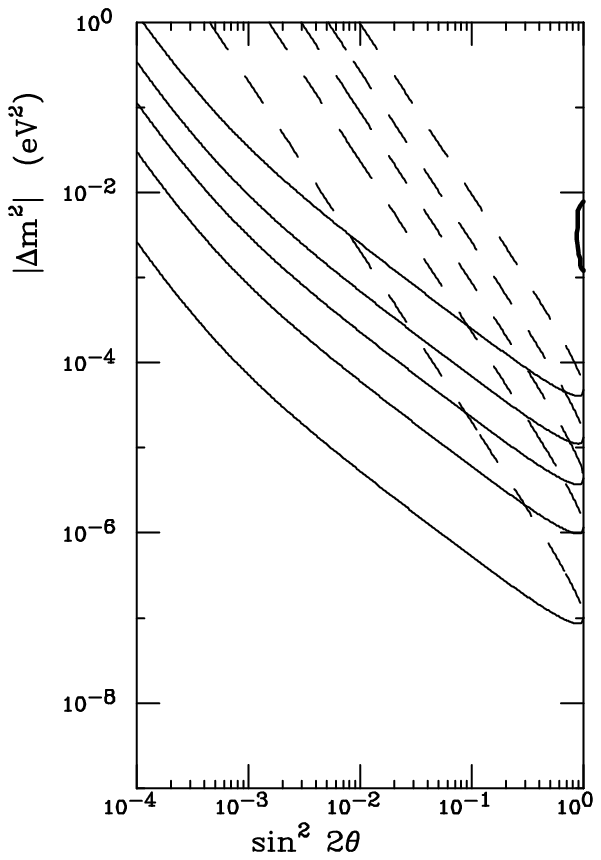}
\hspace{15mm} 
\epsfxsize=11pc
\epsfbox{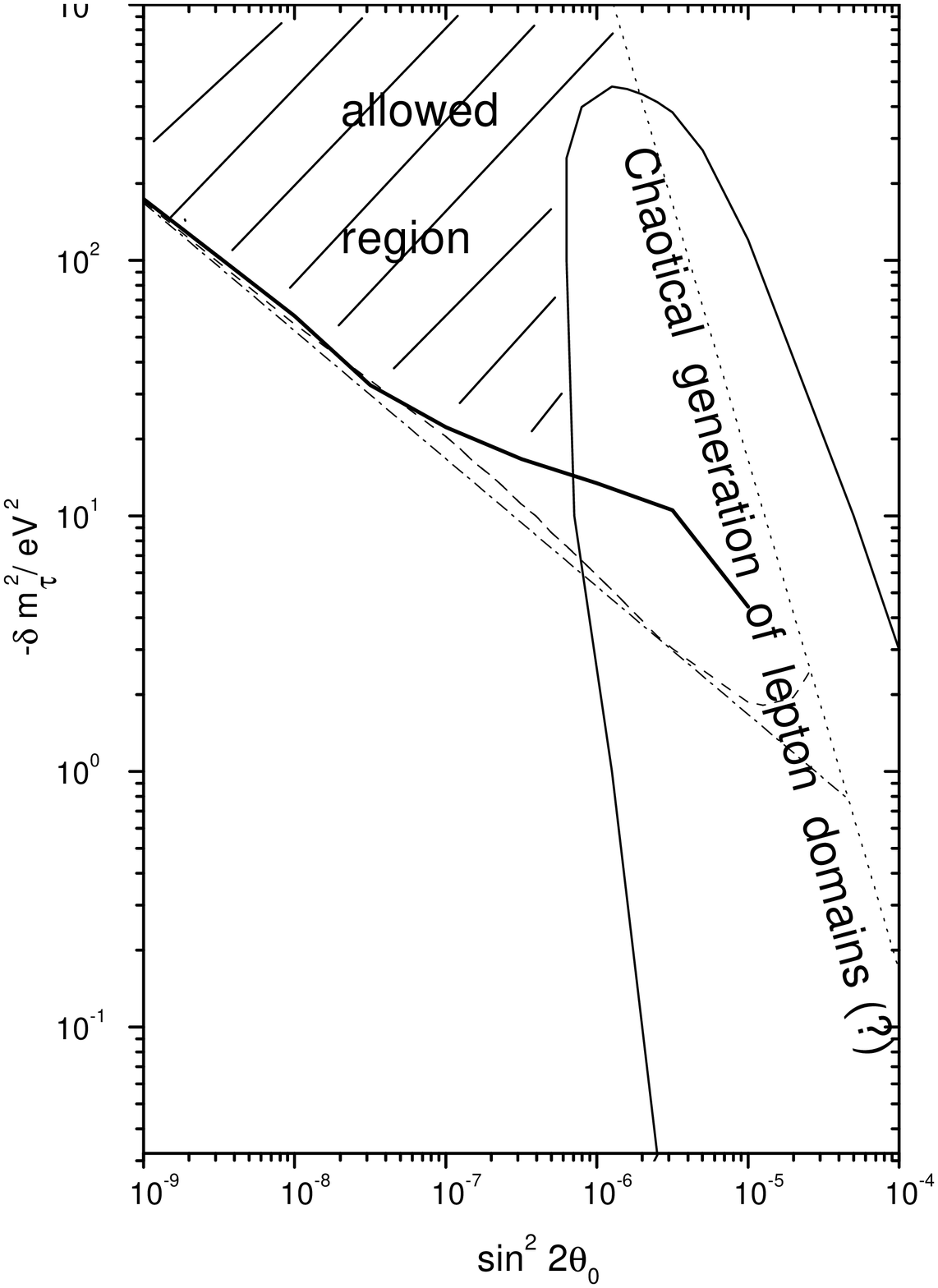} 
\caption{Left: constraints on the mixing parameters for $\nu_{\mu}\leftrightarrow\nu_{s}$.
The dashed lines correspond the non resonant case positive $\delta m^{2}$), 
the thin solid lines to the resonant one (negative $\delta m^{2}$). From bottom up the different lines correspond to $N_{\nu_{s}}^{\rm eff}=0.1, 0.3, 0.5, 0.7, 0.9$; 
the allowed region for the solution of the atmospheric neutrino problem  
is also shown (thick solid line). Right: allowed region in the three neutrino mixing mechanism.}
\vspace{-8mm}
\end{figure}
have been calculated analytically \cite{DiBari99} in the effective kinetic approach
(static approximation) developed by Foot and Volkas \cite{Foot96,Foot97,Bell98}.
They are in a good agreement with the numerical calculations \cite{Enqvist92}.
Moreover they are valid when it is assumed that the 
{\em effective total lepton number} 
$L\equiv 2L_{\nu_{\mu}}+L_{\nu_{\tau}}+L_{\nu_{e}}-(1/2)\,B_{n}$
starts and remains negligible. The quantities 
$Q_{X}\equiv (n_{X}-n_{\bar{X}})/n_{\gamma}$ ($Q=L$ or $B$) are the 
{\em asymmetries} of the particle species $X$.
For positive $\delta m^{2}$ this picture is correct, but for
negative 
\footnote{We define $\delta m^{2}\equiv m^{2}_{2}-m^{2}_{1}$, with 
$m_{1}\,(m_{2})$ the eigenvalue of the mass eigenstate coinciding
with the active (sterile) neutrino interaction eigenstate for zero mixing.}
$\delta m^{2}$, even though one starts from a situation
in which $L$ is initially negligible ($L\ll 10^{-6}$), at a critical
temperature $T_{c}\simeq 18\,{\rm MeV}\,|\delta m^{2}|^{1/6}$ 
this can undergo a phase of rapid growth,
first exponentially and afterwards as a power-law \cite{Foot96,Foot97}.
The presence of a large lepton number suppresses the sterile neutrino 
production \cite{Foot96,DiBari99} and thus it is legitimate to suspect 
whether taking into account the generation of lepton number can relax 
the constraints. The answer is negative. First, the growth occurs only for 
very small mixing angles and thus in anycase the BBN bound for the 
atmospheric neutrino solution cannot be evaded and moreover, even for 
small mixing angles, it occurs too late, when sterile neutrinos have already
mostly been produced: therefore the constraints cannot be significantly relaxed.
The account of lepton number generation is thus uneffective
in a simple two neutrino mixing scenario.

\section{Three neutrino mixing mechanism to evade the BBN bound}

 Assuming that the sterile neutrino is also 
slightly mixed with a heavier tau neutrino,  
a lepton number $L$ can be generated during
$\nu_{\tau}\leftrightarrow\nu_{s}$ oscillations and it 
can afterwards suppress
the sterile neutrino production during the 
$\nu_{\mu}\leftrightarrow\nu_{s}$ oscillations \cite{Foot97}.
Neutrino oscillations are maximally active at 
$T_{c}\propto|\delta m^{2}|^{\frac{1}{6}}$. 
In this case we have two different $T_{c}$, associated 
with the two different $\delta m^{2}$ that we indicate with 
$\delta m^{2}_{\mu}$ and $\delta m^{2}_{\tau}$.
Therefore,
to have a generation of lepton number before the sterile neutrino production,
 one has to impose that $\delta m^{2}_{\tau}>\delta m^{2}_{\mu}$. 
Moreover it is clear that one has also to impose that a significant sterile 
neutrino production does not occur already during 
the oscillations $\nu_{\tau}\leftrightarrow\nu_{s}$. 
Therefore the constraints discussed previously
in the two neutrino mixing scenario
must be imposed on the mixing parameters of tau neutrino
(dotted line in right side of figure 1). 
These conditions are still not sufficient and things are made  
more complicate considering that the two different neutrino oscillations 
are mutually dependent. While the lepton number is produced 
from $\nu_{\tau}\leftrightarrow\nu_{s}$, it has also 
the effect of increasing the temperature $T_{c}$ for  
$\nu_{\mu}\leftrightarrow\nu_{s}$ oscillations that 
are anticipated and participate to the rate of growth of lepton number
but with a destroying contribution. If this counter 
effect is dominant, first the lepton number stops its growth and then 
is completely destroyed. To avoid this situation the condition 
$\delta m^{2}_{\tau}>\delta m^{2}_{\mu}$ must be largely satisfied. 
However structure formation arguments do not allow a 
tau neutrino mass much higher than a few eV's and thus 
the lower limit on the $\delta m^{2}_{\tau}$ must not be 
much larger than about $10\,{\rm eV}^{2}$. This can be determined 
analitically \cite{DiBari99c} in the effective kinetic approach. 
The rate of the total lepton number $L$ is simply given by 
$dL/dt=2dL_{\nu_{\mu}}/dt+dL_{\nu_{\tau}}/dt$. The first term is the contribution 
from $\nu_{\mu}\leftrightarrow\nu_{s}$ and always destroys $|L|$, 
while the second is the contribution from $\nu_{\tau}\leftrightarrow\nu_{s}$ 
that at the critical temperature drive the growth of $|L|$. 
 It is possible to show that $|dL_{\nu_{\alpha}}/dt|=k_{\alpha}\,\sin^{2}\theta_{\alpha}\,|\delta m^{2}_{\alpha}|$
where $k_{\alpha}$ is a function of time and does not depend on the mixing parameters.
The numerical analysis shows \cite{DiBari99} that if the
lepton number stops only once during its growth, then its fate is to be 
destroyed, otherwise it can grow up to a final value able to suppress the
sterile neutrino production. In this way the condition that one has to impose 
for the lepton number to grow is simply that $d|L|/dt>0$ at any time.
This is equivalent to impose the condition
$|\delta m^{2}_{\tau}|>\,\sqrt{C}\,|\delta m^{2}_{\mu}|/\sqrt{s^{2}_{\tau}}$
where $C$ is the maximum of the ratio $k_{\mu}/k_{\tau}$ during the evolution of
the lepton number. 
In the right panel of figure 1 the the dot-dashed line corresponds
to $|\delta m^{2}_{\mu}|=10^{-3}{\rm eV}^{2}$ and $C=28$, 
the value that gives the best fit of the numerical result 
obtained using the static approximation (dashed line). 
The thick solid line is the result of a numerical calculation in which 
the full quantum kinetic equations have been used \cite{f}.

\section{Chaotical generation of lepton domains ?}

The three neutrino mixing mechanism is independent on the final sign of
the lepton number. If however the final sign is sensitive to
small fluctuations, one can imagine that
different points of the early Universe evolve a different 
sign and that a chaotical generation of lepton domains occurs 
\cite{Foot96,Shi99}.
In this case one should calculate a further sterile neutrino
production deriving from those neutrinos that, crossing the boundaries 
of lepton domains, encounter a new resonance. This additive production could 
spoil the evasion of the BBN bound \cite{Shi99}. Is a chaotical generation of 
lepton domains possible ?  A definitive answer to this difficult problem can be obtained
only performing the full quantum kinetic calculations including momentum 
dependence \cite{DiBari99c}. The results show that the sign is fully determined for
a large choice of mixing parameters and only in a restricted region the numerical
analysis cannot be conclusive at the present. This region is the thin 
solid line in the right panel of the figure.
It is evident that even assuming that a chaotical generation of lepton domains
occurs in this region, determining a sterile neutrino overproduction,
the allowed region for the three neutrino mixing mechanism still includes values
of $\delta m^{2}\ll 100 {\rm eV}^{2}$, corresponding to a tau neutrino mass of
a few eV's. Thus the conclusion is that {\em cosmology cannot exclude
the solution $\nu_{\mu}\leftrightarrow\nu_{s}$ to the atmospheric neutrino anomaly}.

\section*{Acknowledgments}
I wish to thank Robert Foot, Paolo Lipari, Maurizio Lusignoli and
Ray Volkas for the collaboration and the encouragement during 
the period of thesis; A.D. Dolgov, K. Enqvist, K. Jedamzik, 
K. Kainulainen, S. Pastor and A. Sorri for nice discussions
during the meeting; the organizers for a great conference.

\end{document}